\documentclass[
reprint,
superscriptaddress,
amsmath,amssymb,
aps,prl
]{revtex4-2}
\usepackage{natbib}
\usepackage{graphicx}
\usepackage{dcolumn}
\usepackage{bm}
\usepackage{xcolor}
\usepackage{soul}
\usepackage[normalem]{ulem}
\sethlcolor{yellow}
\setstcolor{red}

\newcommand\rout{\bgroup\markoverwith
{\textcolor{red}{\rule[.5ex]{2pt}{0.4pt}}}\ULon}
\newcommand{\code}[1]{\texttt{#1}}

\newcommand{\dStar}{\code{dStar}}
\newcommand*{\Qimp}{\ensuremath{Q_\mathrm{imp}}} 

\newcommand*{\Qsh}{\ensuremath{Q_\mathrm{sh}}} 
\newcommand*{\Psh}{\ensuremath{P_\mathrm{sh}}} 
\newcommand*{\Tc}{\ensuremath{T_\mathrm{core}}} 

\newcommand*{\system}[3]{\ensuremath{\mathrm{#1~{#2}{#3}}}}
\newcommand*{\KS}{\system{KS}{1731}{-260}}
\newcommand*{\nuclei}[2]{\ensuremath{\mathrm{^{#1}#2}}}
\newcommand*{\iron}[1][56]{\nuclei{#1}{Fe}}

\newcommand*{\punit}{\ensuremath{\mathrm{dyn\,cm^{-2}}}}
\newcommand*{\rhounit}{\ensuremath{\mathrm{g\,cm^{-3}}}}

\begin{document}


\title{Crust composition and the Shallow Heat Source in KS 1731-260}

\author{R. Jain}
\email{jain15@llnl.gov}
    \affiliation{Nuclear and Chemical Sciences Division, 
    Lawrence Livermore National Laboratory,
    Livermore, CA 94551, USA}
\author{E. F. Brown}
    \affiliation{Facility for Rare Isotope Beams,
    Michigan State University,
    East Lansing MI 48824 USA}
    \affiliation{Department of Physics and Astronomy,
    Michigan State University,
    East Lansing MI 48824 USA}
    \affiliation{Department of Computational Mathematics, Science and Engineering,
    Michigan State University,
    East Lansing MI 48824 USA}
\author{H. Schatz}
    \affiliation{Facility for Rare Isotope Beams,
    Michigan State University,
    East Lansing MI 48824 USA}
    \affiliation{Department of Physics and Astronomy,
    Michigan State University,
    East Lansing MI 48824 USA}
\author{A. V. Afanasjev}
    \affiliation{Department of Physics and Astronomy,
    Mississippi State University,
    Mississippi State, MS 39762, USA}
\author{M. Beard}
    \thanks{Deceased.}
    \affiliation{Department of Physics,
    University of Notre Dame,
    Notre Dame, IN 46556, USA}
\author{L. R. Gasques}
    \affiliation{Universidade de Sao Paulo, Instituto de Fisica, 
    Rua do Matao, 1371, 05508-090, Sao Paulo, SP, Brazil}
\author{J. Grace}
    \affiliation{Department of Physics and Astronomy,
    Michigan State University,
    East Lansing MI 48824 USA}
    \affiliation{Department of Computational Mathematics, Science and Engineering,
    Michigan State University,
    East Lansing MI 48824 USA}
\author{A. Heger}
    \affiliation{School of Physics and Astronomy,
    Monash University, Vic 3800, Australia}
\author{G. W. Hitt}
    \affiliation{Department of Physics and Engineering Science,
    Coastal Carolina University,
    P.O. Box 261954, Conway, SC 29528, USA}
\author{W. R. Hix}
    \affiliation{Physics Division, Oak Ridge National Laboratory, 
    P.O. Box 2008, Oak Ridge, TN 37831-6354, USA}
    \affiliation{Department of Physics and Astronomy,
    University of Tennessee Knoxville,
    Knoxville, TN 37996-1200, USA}
\author{R. Lau}
    \affiliation{ HKU SPACE, PO LEUNG KUK, Stanely Ho Community Colllege,
    Hong Kong}
\author{W.-J. Ong}
    \affiliation{Nuclear and Chemical Sciences Division, 
    Lawrence Livermore National Laboratory,
    Livermore, CA 94551, USA}
\author{M. Wiescher}
    \affiliation{Department of Physics,
    University of Notre Dame,
    Notre Dame, IN 46556, USA}
\author{Y. Xu}
    \affiliation{Extreme Light Infrastructure - Nuclear Physics (ELI-NP), Horia Hulubei National Institute for R\&D in Physics and Nuclear Engineering (IFIN-HH), 30 Reactorului Street, 077125 Magurele, Romania}
    
\date{\today}

\begin{abstract}

The presence of a strong shallow heat source of unknown origin in accreting neutron star crusts has been inferred by analyzing X-ray observations of their cooling in quiescence. We model the cooling of \KS\, using realistic crust compositions and nuclear heating and cooling sources from detailed nuclear reaction network calculations. We find that the required strength of the shallow heat source in \KS\, is reduced by more than a factor of 3 compared to previous analysis, a 5$\sigma$ difference that alleviates the need for exotic solutions. Our analysis also suggests the existence of an impure nuclear pasta layer in the inner crust of \KS\, though future observations will provide more stringent constraints. In addition, we obtain constraints on the dominant surface burning modes of \KS\, over its history.

\end{abstract}

\maketitle

Neutron stars in binary systems can accrete matter from their companion star \cite{Chevalier1993}. Quasi-persistent transiently accreting neutron star systems are a subset of soft X-ray transients that are characterized by extended alternating periods of accretion and quiescence that can last years to decades \cite{Meisel2018,Wijnands2017}. They are bright X-ray sources during an accretion outburst with luminosities of $10^{36-39}$~ergs. During quiescence, this luminosity drops by many orders of magnitude, revealing the thermal emission of the neutron star crust. 
X-ray observations during extended periods of quiescence indicate a decrease in luminosity over a timescale of months. 
This is interpreted as cooling of the neutron star crust that was heated during 
earlier accretion outbursts \cite{Wijnands2017}. The cooling rate depends on the distribution of heat sources, cooling mechanisms, and the thermal transport properties of the crust, and thus carries information about its structure and composition.  Comparison of models of neutron star crust cooling with X-ray observations have led to many new insights, such as the well-ordered lattice structure of the outer crust \citep{Cackett2006Cooling-of-the-,Shternin2007Neutron-star-co,Brown2009Mapping-Crustal}, the presence of neutron superfluidity in the inner crust \citep{Shternin2007Neutron-star-co,Brown2009Mapping-Crustal}, and the possible existence of nuclear pasta at the crust-core transition \citep{Horowitz2015,Deibel2017}. 

During accretion, the neutron star crust is heated by nuclear reactions that are induced by the steadily increasing density of matter
throughout the crust 
and include electron captures, neutron captures, neutron transfer reactions, $\beta$-decays, and pycnonuclear fusion reactions \cite{Haensel2008,Lau2018,Schatz2022,Shchechilin2022,Shchechilin2023}. However, to match the X-ray observational data at early cooling times in most sources, models have to include an additional, relatively shallow heat source of unknown origin \citep{Brown2009Mapping-Crustal,Degenaar2014Probing-the-Cru,Turlione2015,Waterhouse2016,Potekhin2021}. Additional evidence for a hotter-than-expected crust during accretion comes from observations of superbursts, thermonuclear explosions of an accumulated layer of carbon \cite{kuulkers.ea:ks1731-superburst}.  Reconciling observed superburst recurrence times of the order of years and inferred ignition depths with models requires an additional shallow heat source of similar magnitude that facilitates superburst ignition \cite{Meisel2024}. 

Many theories for the origin of this extra heating have been proposed, for example, viscous heating caused by accretion induced shear \citep{Piro2007Turbulent-Mixin}, convective mixing \citep{Horowitz2007Phase-Separatio}, nuclear fusion of light neutron-rich elements \citep{Horowitz2008Fusion-of-neutr}, electron captures in the crust \citep{Gupta2006Heating-in-the-,Chamel2020}, pion-induced heating \citep{Fattoyev2018}, gravity-wave transport of angular momentum \citep{Inogamov2010}, or hyperbursts powered by explosive burning of an O-Ne mixture \citep{Page2022A-Hyperburst-in}. The strength and location of the additional unknown shallow heating can be constrained by matching models with the observed cooling of accreting neutron stars in quiescence. The inferred heating varies from a few MeV per accreted nucleon (MeV/u) for most transient systems to more than 10 MeV/u for \system{MAXI}{J0556}{-332} \citep{Deibel2015A-Strong-Shallo,Page2022A-Hyperburst-in}. Limited data from multiple observed outbursts of the same source indicate that in some systems such as \system{MXB}{1659}{-29}, the shallow heating per accreted nucleon remains the same \citep{Parikh2019}, whereas in other systems, such as \system{Aquila}{X}{-1}, it varies from outburst to outburst \citep{Degenaar2019}. 
All these inferences about shallow heating depend on accurate modeling of the thermal conductivity, which, in the crust, is largely determined by impurity scattering of electrons and is characterized by the impurity parameter \Qimp\, \citep{Brown2009Mapping-Crustal} \citep{itoh93}. 
The impurity parameter is defined as \Qimp\, = $\sum_{i} Y_i (Z_i - \langle Z \rangle)/\sum_i Y_i$ with isotopic abundances $Y_i$, element number $Z_i$, and average element number $\langle Z \rangle$. 

Cooling curve fits for quasi-persistent transients typically require \Qimp\, $<$ 10 \cite{Brown2009Mapping-Crustal,Merritt2016The-Thermal-Sta} with models either implementing an average \Qimp\, for the entire crust, or separate average parameters for outer and inner crusts. Models, however, predict considerable variations of \Qimp\, as a function of depth \cite{Lau2018}. Hydrogen and helium burning on the neutron star surface, either in form of Type I X-ray bursts, or as stable burning, produce a broad range of elements up to $Z \approx 50$ \cite{Woosley2004,Keek2012} resulting in a \Qimp\, of up to 100 in the surface layers. \citet{Lau2018} used a nuclear reaction network to track the evolution of  \Qimp\, as a function of depth as nuclear reactions modify the composition and heat the crust. While they found a gradual decrease of \Qimp\, with depth, significant impurities are expected to persist in much of the outer crust, depending on the initial composition created by thermonuclear burning on the surface. Here we implement the detailed \Qimp\, profiles and associated nuclear energy generation profiles for the outer crust obtained with the nuclear reaction network described in \cite{Lau2018} into a crust cooling code and show that this affects the constraints of system parameters such as the properties of additional shallow heating significantly. We also reexamine the possible role of a high impurity nuclear pasta layer at the bottom of the inner crust \cite{Horowitz2015, Deibel2017}. 

We apply our updated model to the quasi-persistent transient system \KS, which 
entered quiescence in early 2001 \citep{wijnands:ks1731} after a $\sim$ 12.5 year outburst phase \citep{Syunyaev1990The-New-X-Ray-T}. Since then, a number of observations with \emph{Chandra} and \emph{XMM-Newton} \citep{wijnands:ks1731,wijnands.ea:xmm_1731,Cackett2006Cooling-of-the-,Cackett2010Continued-Cooli,Merritt2016The-Thermal-Sta} have tracked the cooling of \KS. The data indicate that around 3,000 days into quiescence the X-ray luminosity leveled out, which has been interpreted as the crust having cooled to the temperature of the core. The previous model fits to the cooling light curve of \KS\, have obtained a shallow heat source with a strength of 1.36 $\pm$ 0.18 MeV/u, a core temperature of 9.35 $\pm$ 0.25 $\times$ $10^7$ K, and an average \Qimp\, for the crust 
of $4.4_{-0.5}^{+2.2}$ \cite{Merritt2016The-Thermal-Sta}. 

\begin{figure}[tbp]
\includegraphics[width=\linewidth]{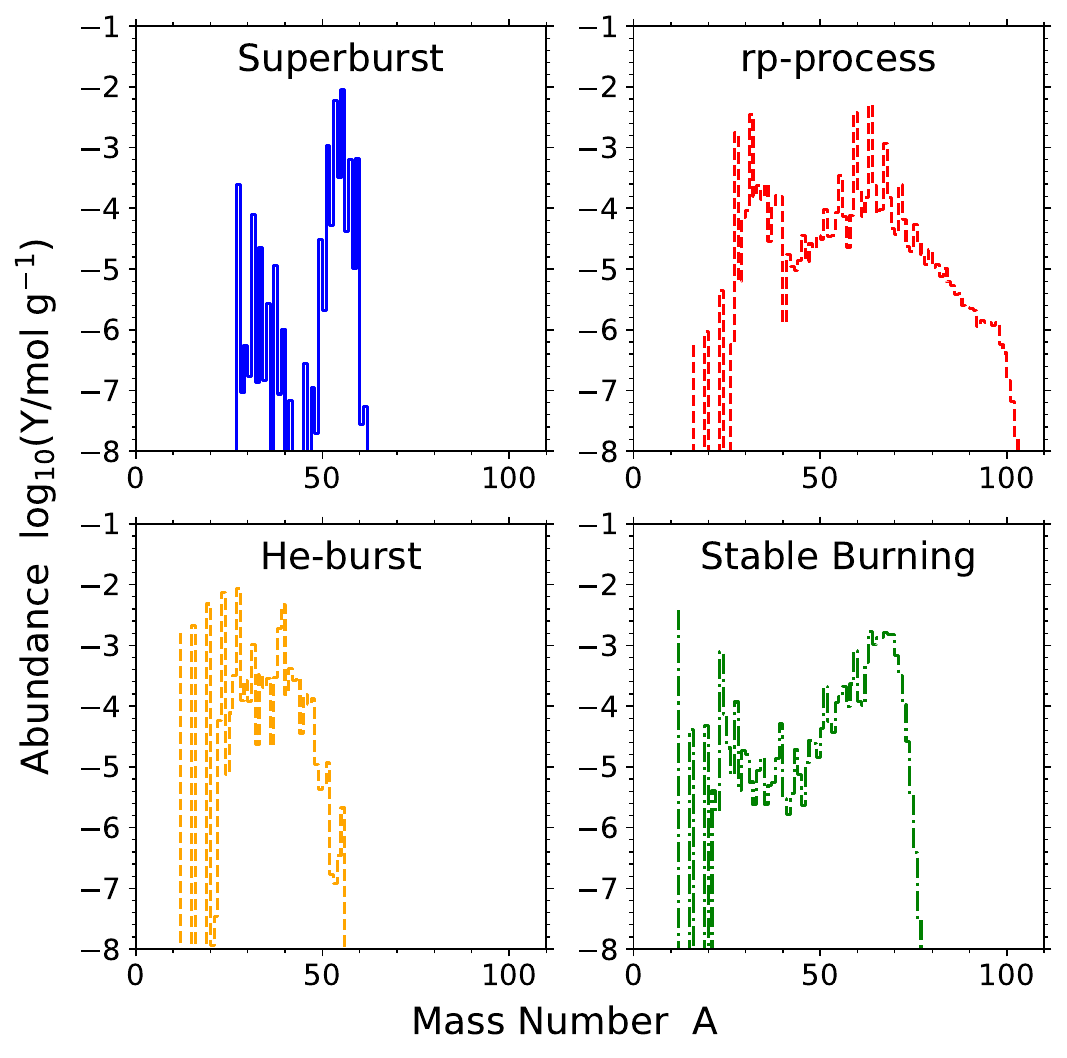}
\caption{The abundances as a function of mass number for the ashes of superbursts (blue) \citep{Keek2012}, mixed H/He bursts undergoing rp-process (red) \citep{Cyburt2016}, He powered X-ray bursts (orange) \citep{Woosley2004}, and 
stable nuclear burning (green) \citep{Schatz1999}.
\label{fig:abd}}
\end{figure}

The starting point for a calculation of the crust nuclear reactions is the composition of the ashes from thermonuclear burning during the accretion outburst. In principle, X-ray observations during the outburst phase can constrain this. \KS, however, exhibits a wide range of burning behaviors.
As the complete burning history 
over the roughly 10,000 years of outburst that define the composition of the current crust is not known, it is unclear which of these burning modes have produced the bulk of the current crust. We perform our analysis with a range of realistic ash compositions that represent the various observed modes of thermonuclear burning. 
A total of 366 thermonuclear X-ray bursts have been detected from \KS\, with various instruments \citep{Galloway2020The-Multi-INstr}. These bursts include short bursts with timescales of seconds during phases of high accretion (`banana branch') that indicate flashes of He burning. This burning regime has $\alpha \sim$  200--690, where $\alpha$ is the ratio of energy released in X-rays during fuel accretion to energy released in bursts. This indicates the presence of significant concurrent stable nuclear burning. During the phases of low accretion rate (`island state'), longer X-ray bursts with durations of 30 s are observed, indicating mixed hydrogen and helium burning via the $\alpha$-p and rapid proton capture (rp) processes. A superburst, thought to be powered by explosive burning of an accumulated carbon layer, has also been observed from \KS\, \cite{kuulkers.ea:ks1731-superburst}. Each of these burning modes produces a distinct abundance distribution (Fig.~\ref{fig:abd}). We use the results from \code{KEPLER} \citep{WZW78} for the ashes of He powered X-ray bursts \citep{Woosley2004} (accretion rates of $3.5 \times 10^{-10}\,\mathrm{M}_{\odot}\,\mathrm{yr}^{-1}$ with solar metallicity), mixed H/He bursts (rp-process) \citep{Cyburt2016} and superbursts \citep{Keek2012}.
For stable burning we adopt the steady state burning results from \citet{Schatz1999} for an accretion rate of $10^{-8}\,\mathrm{M}_{\odot}\,\mathrm{yr}^{-1}$.
For comparison with previous work we also perform calculations assuming thermonuclear burning produces pure $^{56}$Fe.

For each initial composition at the top of the crust, we calculate as a function of depth the composition, 
\Qimp, nuclear heating, and neutrino cooling via crust Urca cooling \cite{Schatz2013, Deibel2016} using the nuclear reaction network \code{XNet} \cite{Hix1999}. 
\code{XNet} includes a comprehensive set of nuclear reactions such as electron captures, $\beta^{-}$ decays, neutron capture/emission, neutron transfer, and nuclear fusion (see \citep{Lau2018, Jain2023, Schatz2022} for details). 

\begin{figure}[tbp]
\includegraphics[width=\linewidth]{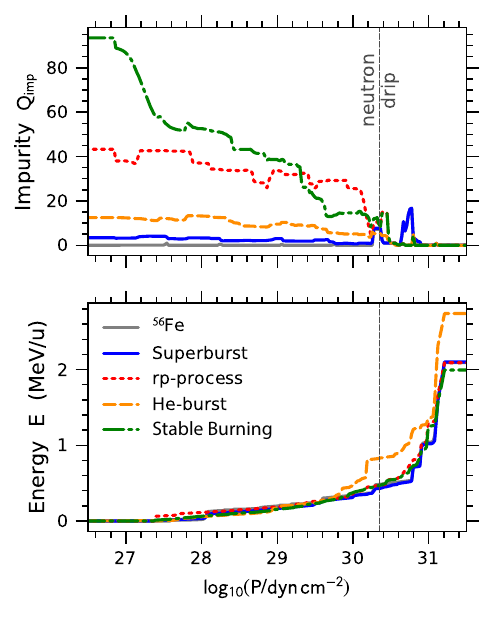}
\caption{The upper panel shows the impurity parameter profile and the lower panel shows the nuclear heating profile in the neutron star crust. Nuclear heating is plotted as the total integrated energy deposited up to the given depth. Different curves correspond to different crust compositions considered in this study. 
\label{fig:xnet}}
\end{figure}

The top panel in Figure \ref{fig:xnet} shows the resulting impurity parameter profiles for the different initial compositions considered in this study. The crusts composed of stable burning ashes have the highest impurity, followed by rp-process ashes and then by He-burst ashes. The crusts composed of pure \iron\, or superburst ashes are the most pure. There is a significant increase in impurity at depths of about log(P) $\sim$ 30.6 -- 30.8 for these compositions. This is attributed to the onset of superthreshold electron-capture cascades (SEC) and SEC-pycnonuclear fusion cycles as discussed in Lau \emph{et al.} \citep{Lau2018}. 
The bottom panel in Figure \ref{fig:xnet} shows the resulting nuclear heating profiles in the crust. 
All crust compositions lead to a similar heating profiles with the exception of He-burst ashes. Since He-bursts are characterized by incomplete burning of nuclear matter on the surface, their ashes carry less binding energy per nucleon on average and hence deposit more energy in the crust. 

The nuclear reaction network outputs are then implemented into the neutron star thermal transport code \dStar\, \cite{Brown2015}. \dStar\, is a flexible, modular software package 
for calculations of the thermal evolution of the neutron star crust during accretion outbursts and subsequent quiescence which takes into account a variety of heat sources and cooling mechanisms. \Qimp\, and nuclear heating profiles are directly taken from the network calculations, while Urca cooling layers are implemented as temperature dependent, neutrino cooling layers, with the strength and location obtained from the nuclear network calculations. 

Nuclear theory predictions indicate the possibility of a nuclear pasta layer in the deepest layers of the inner crust with nucleonic densities greater than 0.05 nucleons/fm$^{3}$ \citep{OYAMATSU1993431,Caplan2017}. Molecular dynamics simulations predict significant impurities in the pasta layer resulting in a relatively high \Qimp = 40, which has been shown to significantly affect models predicting the cooling of \KS\, and other systems \cite{Horowitz2015,Deibel2017}. We therefore carry out calculations with and without such a pasta layer located at a pressure beyond 3.5$\times 10^{32}$  \punit, corresponding to nuclear pasta density of 9$\times 10^{13}$ \rhounit. 

 Our model of \KS\, accretes at a constant rate of 10$^{17}$ g s$^{-1}$ \citep{Galloway2008Thermonuclear-t} for 4,383\,days before transitioning into quiescence. The system is allowed to evolve over time and the redshifted effective surface temperature ($\mathrm{T_{eff}^{\infty}}$) is tracked at regular intervals to get the modeled cooling curves. Technical details on the modeling of \KS\, can be found in the Supplemental material \cite{supp} (see also references \cite{Akmal1998,Timmes2000,Chabrier1998,Farouki1993,potekhin97,baiko98:_ion_coulom,Levenfish1994,Gandolfi2008,yakovlev1998neutrinoemissioncooperpairing,Steiner2009,compositions} therein).

The model results are compared to quiescent X-ray observations of \KS\, with \emph{Chandra} and \emph{XMM-Newton} satellites carried out over the years \citep{wijnands:ks1731,wijnands.ea:xmm_1731,Cackett2006Cooling-of-the-,Cackett2010Continued-Cooli,Merritt2016The-Thermal-Sta}. 
\begin{figure}[tbp]
\includegraphics[width=\linewidth]{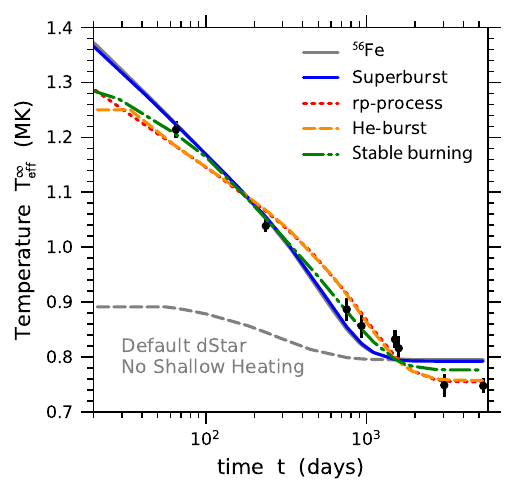}
\caption{The best-fit cooling curves for \KS\, for different outer crust compositions without a nuclear pasta layer. The best-fit values for the shallow heating parameters are listed in Table \ref{table:cc}.  
\label{fig:cc_nopasta}}
\end{figure}
As pointed out in previous work, the calculations based on known nuclear heating sources are unable to reproduce the observations at early cooling times (Fig.~\ref{fig:cc_nopasta}). Consequently, an artificial shallow heat source was implemented in \dStar\, with its strength \Qsh\, and location at pressure \Psh\, being free parameters. The core temperature \Tc\, is also chosen to be a free parameter that acts as a boundary condition for heat transfer. The three free parameters \Qsh, \Psh, and \Tc\, are then determined from the observational data using Bayesian inference. We start with wide uniform priors for all three parameters and consider Gaussian likelihoods. Posterior distributions are sampled using MCMC sampling. The median of the distributions is chosen to be the best-fit value and asymmetric errors are quoted at the 68\% credibility intervals ($16^{\mathrm{th}}$ and $84^{\mathrm{th}}$ percentile). Figures \ref{fig:cc_nopasta} and \ref{fig:cc_pasta} show the best-fit cooling curves for different crust compositions of \KS\, without and with a nuclear pasta layer in the inner crust, respectively. Table \ref{table:cc} lists the corresponding best-fit values for the shallow heating parameters. One finds that different crust compositions require very different levels of shallow heating to fit the same observational data; crusts composed of \iron\, 
or superburst ashes require more than 3 times the shallow heating 
than crusts composed of stable burning ashes. In contrast, shallow heating depth and core temperature are relatively robust and do not depend strongly on the composition of the outer crust.

\begin{table*}
\caption{Best-fit parameters for the shallow heat source in \KS\, from crust cooling observations in quiescence for different outer crust compositions. They are reported for models with and without a nuclear pasta layer at the crust-core transition.
\label{table:cc}}
\begin{tabular}{ c | c c c c c c c c c}
\hline \hline
Nuclear Pasta & Crust & & Strength & & Depth & & Core Temperature & & reduced--$\chi^{2}$\\
Layer & Composition & & \Qsh (MeV/u) & & $\log_{10}(\Psh/\punit)$ & & $(\Tc/\mathrm{10^{7}\,K})$ & & (8 -- 3 = 5) dof\\
\hline
& \iron\, ashes & & $1.60_{-0.10}^{+0.12}$ & & $27.2_{-0.8}^{+0.8}$ & & $7.02_{-0.14}^{+0.14}$ & & 5.84 \\
& Superburst ashes & & $1.52_{-0.10}^{+0.11}$ & & $27.1_{-0.8}^{+0.9}$ & & $6.98_{-0.13}^{+0.14}$ & & 5.47 \\
No & rp-process ashes & & $0.61_{-0.05}^{+0.05}$ & & $26.6_{-0.4}^{+0.6}$ & & $6.55_{-0.15}^{+0.15}$ & & 3.37 \\
& He-burst ashes & & $0.82_{-0.06}^{+0.06}$ & & $26.7_{-0.4}^{+0.5}$ & & $6.40_{-0.14}^{+0.15}$ & & 3.66 \\
& Stable burning ashes & & $0.45_{-0.03}^{+0.04}$ & & $26.9_{-0.6}^{+0.5}$ & & $7.09_{-0.17}^{+0.26}$ & & 2.75 \\
\hline
& \iron\, ashes & & $1.23_{-0.08}^{+0.09}$ & & $26.7_{-0.5}^{+0.7}$ & & $5.77_{-0.16}^{+0.16}$ & & 1.39 \\
& Superburst ashes & & $1.17_{-0.08}^{+0.09}$ & & $26.7_{-0.5}^{+0.7}$ & & $5.77_{-0.17}^{+0.17}$ & & 1.63 \\
Yes & rp-process ashes & & $0.46_{-0.04}^{+0.04}$ & & $26.6_{-0.4}^{+0.6}$ & & $5.73_{-0.17}^{+0.18}$ & & 6.75 \\
& He-burst ashes & & $0.52_{-0.04}^{+0.05}$ & & $26.7_{-0.1}^{+0.1}$ & & $5.31_{-0.17}^{+0.16}$ & & 8.89 \\
& Stable burning ashes & & $0.38_{-0.03}^{+0.03}$ & & $26.7_{-0.4}^{+0.4}$ & & $6.19_{-0.17}^{+0.17}$ & & 1.52 \\
\hline
\end{tabular}
\end{table*}

\begin{figure}[tbp]
\includegraphics[width=\linewidth]{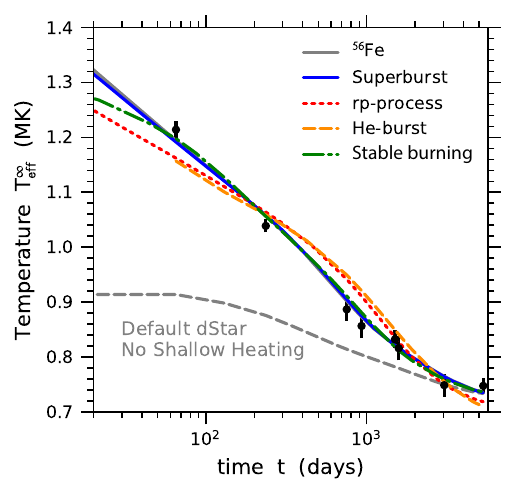}
\caption{The same as in Fig. \ref{fig:cc_nopasta} but for crust compositions which include a nuclear pasta layer at the crust-core transition.
\label{fig:cc_pasta}}
\end{figure}

We find significant differences in the fit quality and the best fit \Qsh\, for different composition and heating profiles in the outer crust. This highlights the importance of the nuclear reactions during thermonuclear burning on the surface, and the importance of tracking the detailed composition in the outer crust. In principle, this can be used to constrain the outer crust composition and thus to define the dominant thermonuclear burning mode of the system. However, the conclusions depend significantly on whether a nuclear pasta layer is included or not. Without a nuclear pasta layer, stable burning ashes are preferred while relatively pure superburst and pure \iron\, ashes are significantly disfavored. This suggests that stable burning in the high accretion rate state is the dominant thermonuclear surface burning mode of \KS. This burning mode does produce significant amounts of carbon and is one of the most promising pathways to explain the existence of superbursts since unstable H/He burning tends to produce only negligible amounts of carbon. The disfavoring of superburst ashes would then imply that the conditions are such that the carbon produced in stable burning is itself burned mostly stably, with superbursts being a rare occurrence over the lifetime of the system. However, when a nuclear pasta layer is included, the picture changes. While stable burning ashes still provide a very good fit, the fit quality with a high purity composition such as superburst ashes or $^{56}$Fe is similar. As a consequence, high accretion rate stable burning is again preferred (by producing an outer crust composition that fits observations, or by producing the fuel for superbursts, which in turn produce an acceptable outer crust composition), but there would be no constraint on the frequency of superbursts and the fraction of stable burning ashes they process.

An outer crust composed of stable burning ashes results in an excellent fit of the observational data with a shallow heating strength of just 0.38 MeV/u or 0.45 MeV/u, with and without pasta layer, respectively. These values are considerably lower compared to previous work, e.g. the inferred 1.36 $\pm$ 0.18 MeV/u in the analysis of \citet{Merritt2016The-Thermal-Sta} using a uniform, low \Qimp\, and standard heating profiles. The high impurity in the outer crust slows down the cooling and therefore the crust need not be as hot at the beginning of quiescence to fit the first observation. This new, much lower inferred value for the strength of shallow heating is comparable to the uncertainties in nuclear heating \cite{Chamel2020}, alleviating the need for more exotic solutions to the shallow heating problem, though the location of heat deposition has to be considered as well. The low shallow heating does lead to distinct cooling profiles prior to the first data point of \KS\, 65 days after the onset of quiescence, highlighting the importance of early observations to constrain shallow heating and, according to our new calculations, also the composition of the outer crust. 

Overall we find significantly better fits of the observational data with the models that include a nuclear pasta layer. This confirms the result of \citet{Deibel2017} that was based on a simplified outer crust composition. However, it contradicts previous work that found cooling curves could be fit equally well with and without pasta \cite{Horowitz2015,Merritt2016The-Thermal-Sta}. Inclusion of the pasta layer brings an important change in the interpretation of the leveling off of the observed cooling curve as it does not require reaching the core temperature. Instead, it merely indicates the slow down in cooling produced by the low thermal conductivity pasta layer, and cooling is predicted to continue to a significantly lower core temperature. This has been found before \cite{Horowitz2015, Deibel2017} and in fact has been used to explain the initially surprising continued cooling observed in \system{MXB}{1659}{-29}. Clearly future observations of \KS\, could be used to provide more evidence for the existence of a high impurity pasta layer. One difference in our analysis is the somewhat lower inferred core temperature (\Tc = 5-7 $\times 10^{7}$K, vs 9 $\times 10^{7}$K \citep{Deibel2017}) that would result in stronger continued cooling that may be more straightforward to observe. 

In conclusion, we find that taking into account the detailed composition of the outer crust, defined by the thermonuclear burning on the surface and nuclear reactions in the crust is important for the interpretation of observations of cooling in quasi-persistent soft X-ray transients. The outer crust composition not only defines nuclear heating profiles, as has been pointed out before, but also affects thermal conductivity as a function of depth. Applying our model to \KS, we find that for the best fit realistic outer crust composition, only 0.35-0.45 MeV/u of additional shallow heating is required. This is lower by 5$\sigma$ compared to previous work that obtained values in excess of 1 MeV/u. Such a low value could possibly be reconciled with nuclear heating uncertainties, and thus may alleviate the need for more exotic explanations. It remains to be seen if this result is applicable to other sources, in particular sources where observational data are available at earlier times in the quiescence phase. 

We also find that based on the observed X-ray cooling data during quiescence, the crust is likely to be dominated by the ashes of stable H/He burning, possibly modified further by subsequent superbursts. Our fits to the observational data have some preference for the existence of a low conductivity disordered nuclear pasta layer in the inner crust. Future observations of \KS\, can be used to test this hypothesis more conclusively, and our results predict a stronger signature compared to previous calculations.

Our assumption of the outer crust of \KS\, being dominated by a single thermonuclear burning mode and associated crustal reactions is a simplification. Given the variety of thermonuclear burning modes observed in \KS, 
the crust could in principle be composed of layers with different compositions. Shorter term variations over timescales of years may be affected by sedimentation and phase separation in the liquid ocean \cite{Horowitz2007Phase-Separatio}. This may lead to homogenization or further separation in steady state and remains to be explored. Lateral composition differences from anisotropies in accretion or nuclear burning may also exist \cite{Ushomirsky2002,Morales2022}. We also assume a constant accretion rate during outburst in line with previous work \cite{Galloway2008Thermonuclear-t,Brown2009Mapping-Crustal,Merritt2016The-Thermal-Sta}. \citet{Ootes2016Neutron-star-cr} pointed out that the variations in the accretion rate may affect the modeled cooling curves, mainly prior to the first observed data point in \KS. We performed test calculations with their time variable accretion rate which changed the extracted shallow heat strength by less than 10\%. 

 It was recently suggested \citep{Gusakov2020,Gusakov2021,Shchechilin2023} that accounting for the diffusion of the superfluid neutrons and associated hydrostatic equilibrium conditions induce an early transition to a composition close to equilibrium. This would prevent pycnonuclear fusion in the inner crust and reduce deep nuclear heating. Recent work by \citet{Potekhin2025} demonstrated that cooling curves can be fitted equally well with this model, while still requiring additional ingredients like shallow heating, and variations in thermal conductivity. While the impact of adopting an alternative inner crust model on our results remains to be explored, we do not expect a major impact on our main conclusions in regards to shallow heating, which are primarily based on a novel treatment of the outer crust. 
 
\acknowledgments
This work was supported by the National Science Foundation under Award Nos. PHY-1430152 (JINA Center for the Evolution of the Elements), OISE-1927130 (IReNA), PHY-1913554, and PHY-2209429, and by the U.S. Department of Energy, Office of Science, Office of Nuclear Physics under Award No. DE-SC0013037. E.F.B. acknowledges support under grant 80NSSC20K0503 from NASA. This was work performed under the auspices of the U.S. Department of Energy by Lawrence Livermore National Laboratory under Contract DE-AC52-07NA27344.

\bibliography{main}

\end{document}